\documentclass[12pt]{article}
\usepackage{geometry}
\usepackage{a4}
\usepackage{graphics}
\usepackage{graphicx,epsfig,subfigure}
\usepackage{epsf}
\usepackage{amsmath,amssymb}
\usepackage{caption}
\usepackage{epstopdf}
\usepackage{enumerate}
\usepackage{cite}

\def\be{\begin{equation}}
\def\ee{\end{equation}}
\def\bea{\begin{eqnarray}}
\def\eea{\end{eqnarray}}
\def\pt{\partial}

\begin{document}

\begin{titlepage}

\title{Thermodynamics and phase transition of charged AdS black holes with a global monopole}
\author{}


\date{
Gao-Ming Deng$^{1,2}$\thanks{\noindent gmd2014@emails.bjut.edu.cn}, \, Jinbo Fan$^{1}$, Xinfei Li$^{1}$, Yong-Chang Huang$^{1,3}$
\vskip0.5cm
{\sl $^1$ Institute of theoretical physics, \\ Beijing University of Technology, Beijing 100124, China \\
     $^2$ School of physics and space science, \\ China West Normal University, Nanchong 637002, China \\
     $^3$ CCAST(World Lab.), P.O.Box 8730, Beijing 100080, China}}

\maketitle

\abstract{\noindent
Thermodynamical properties of charged AdS black holes with a global monopole still remain obscure. In this paper, we investigate the thermodynamics and phase transition of the black holes in the extended phase space. It is shown that thermodynamical quantities of the black holes exhibit an interesting dependence on the internal global monopole, and they perfectly satisfy both the first law of thermodynamics and Smarr relation. Furthermore, analysis of the local and the global thermodynamical stability manifests that the charged AdS black hole undergoes an elegant phase transition at critical point. Of special interest, critical behaviors of the black holes resemble a Van der Waals liquid-gas system. Our results not only reveal the effect of a global monopole on thermodynamics of AdS black holes, but also further support that Van der Waals-like behavior of the black holes is a universal phenomenon.
}
\end{titlepage}

\section{Introduction} \label{section1}

In recent years, the study of topological defects such as cosmic strings, domain walls, monopoles and textures has captured considerable attention and still remains one of the most active fields in modern physics. This is mainly due to their fascinating properties which are hoped to give rise to a rich variety of quite unusual physical phenomena in comparison with those of more familiar systems. As an intriguing one, a global monopole \cite{Barriola:1989hx} can be generated during the phase transition of a system composed of a self-coupling triplet scalar field whose original global $O(3)$ symmetry is spontaneously broken to $U(1)$. One can read Ref.\cite{Vilenkin:1994cs} for a nice review. It is worth mentioning that, in 1989, Barriola and Vilenkin \cite{Barriola:1989hx} first discovered the gravitational field of a global monopole and determined the metric of the black hole with a global monopole. Since then, the physical properties of black holes incorporating global monopoles have been studied extensively \cite{Jing:1993np,Yu:1994fy,Dadhich:1997mh,Li:2002ku,Gao:2002hf,Jiang:2005xb,Brihaye:2005qr,Han:2005qy,Wu:2007sw,Chen:2009vz,
Lin:2011zzd,Man:2013sf,Sharif:2015kna,Ahmed:2016ucs,Jusufi:2017lsl,Li:2017kze,Carames:2017ngt}. However, thermodynamical properties of the black holes with global monopoles in AdS spacetime still remain obscure, although it deserves a deeper investigation of the thermodynamics. Motivated by this fact, in this paper, we first focus our attention on investigating thermodynamics of the charged AdS black hole with a global monopole.

Another motivation to explore the charged AdS black holes with a global monopole comes from the pioneering work \cite{Hawking:1982dh} for AdS black hole phase transitions. In Ref.\cite{Hawking:1982dh}, Hawking and Page described a first-order phase transition between Schwarzschild AdS black holes and the thermal AdS space, well known as the Hawking-Page phase transition. Another landmark owes to Chamblin et al. \cite{Chamblin:1999tk,Chamblin:1999hg} who discovered the small-large black hole phase transitions of Reissner-Nordstr\"{o}m AdS black holes and first pointed out that the transition is analogous to a Van der Waals liquid-gas system. Recently, in an in-depth analogy, Kubiznak and Mann \cite{Kubiznak:2012wp} have explicitly verified the coincidence of the black hole critical behaviors with those of the Van der Waals liquid-gas system. Interestingly, analysis in Ref.\cite{Kubiznak:2012wp} involves a prevailing proposal \cite{Kastor:2009wy,Cvetic:2010jb,Dolan:2012jh,Caceres:2015vsa} which states that the cosmology constant $\Lambda$ and its conjugate quantity should be identified as thermodynamic variables, namely, pressure $P$ and volume $V$ respectively, and included in the first law of black hole thermodynamics. Nowadays, this constructive approach has been widely believed to be much more physically sound. On the one hand, a theory would be more fundamental and physical if the constants such as Yukawa couplings, gauge coupling constants, or the cosmological constant arise as vacuum expectation values rather than being fixed a {\it priori}. On the other hand, more pragmatically, in the presence of a cosmological constant, the traditional first law of black hole thermodynamics becomes inconsistent with the Smarr relation. Luckily, the urgent inconsistency can be reconciled if the first law is modified by including the variation of $\Lambda$. Thereafter, based upon the constructive idea, increasing interest has been stirred to study various black holes \cite{Gunasekaran:2012dq,Belhaj:2012bg,Cai:2013qga,Zhao:2013oza,Ma:2013aqa,Zou:2013owa,Dehghani:2014caa,Xu:2014kwa,Johnson:2014pwa,Wei:2014hba,Wei:2015iwa,Hendi:2014kha,
Guo:2015waa,Mo:2016sel,Sherkatghanad:2014hda,Fernando:2016sps,Hendi:2016njy,Pradhan:2016dun,Zeng:2016aly,Poshteh:2016rwc,Bhattacharya:2017hfj}. In a certain sense, the highlight of our work not only contributes to better understanding black hole thermodynamics, but also provides us with important clues to the underlying structure of puzzling quantum gravity.

This paper is outlined as follows. In Sec.\ref{section2}, we concentrate on working out thermodynamical quantities of the charged AdS black hole with a global monopole and checking both the first law of black hole thermodynamics and Smarr relation. Sec.\ref{section3} is devoted to investigating the phase transition and critical behaviors of the black hole in the canonical ensemble. Conclusions will be drawn in Sec.\ref{section4}.

\section{Black hole solution and the thermodynamics} \label{section2}

Barriola and Vilenkin \cite{Barriola:1989hx} discussed the gravitational field of a global monopole resulting from global $O(3)$ symmetry breaking. The simplest model for giving rise to such global monopoles is described by the following Lagrangian
\be
\mathcal{L}_{g m}=\frac{1}{2} \pt_{\mu} \phi^a \pt^{\mu} \phi^a - \frac{\lambda}{4} {\left( \phi^a \phi^a - \eta_0^2\right)}^2, \label{eq01}
\ee
where $\phi^a$ is a triplet of scalar field, $\lambda$ the coupling constant and $\eta_0$ the parameter related to the symmetry breaking scale. When a charged AdS black hole swallows a global monopole, the general static spherically symmetric metric and gauged potential can be given by \cite{Barriola:1989hx}
\be
d\tilde{s}^2=-\tilde{f}(\tilde{r}) d \tilde{t}^2+\frac{1}{\tilde{h}(\tilde{r})} d \tilde{r}^2 +\tilde{r}^2 \left(d \theta^2 + sin^2\theta d\phi^2\right)\,, \label{eq02}
\ee
\be
\tilde{A}=\frac{\tilde{q}}{\tilde{r}}d\tilde{t}\,,\quad
\tilde{f}(\tilde{r})=\tilde{h}(\tilde{r})=1 - 8 \pi \eta_0^2-\frac{2\tilde{m}}{\tilde{r}}+\frac{\tilde{q}^2}{\tilde{r}^2}+\frac{\tilde{r}^2}{l^2}\,, \label{eq03}
\ee
in which $\tilde{m}$ and $\tilde{q}$ are, respectively, the mass parameter and electric charge parameter, $l$ the AdS radius related with the cosmology constant $\Lambda=-3/l^2$. In order to investigate the general properties of the black holes, by introducing the following coordinate transformation
\be
\tilde{t}=\left(1-8\pi \eta_0^2\right)^{-1/2} t\,,\quad \tilde{r}=\left(1-8\pi \eta_0^2\right)^{1/2} r\,, \label{eq04}
\ee
and defining new parameters
\be
m=(1-8\pi \eta_0^2)^{-3/2} \tilde{m}\,,\quad q=(1-8\pi \eta_0^2)^{-1} \tilde{q}\,,\quad \eta^2=8\pi \eta_0^2\,, \label{eq05}
\ee
we can rewrite the line element (\ref{eq02}) as
\be
ds^2=-f(r)dt^2+\frac{1}{h(r)} dr^2 +\left(1-\eta ^2\right) r^2 \left(d\theta^2 + sin^2\theta d\phi^2\right)\,, \label{eq06}
\ee
\be
A=\frac{q}{r}dt\,,\quad
f(r)=h(r)=1-\frac{2 m}{r}+\frac{q^2}{r^2}+\frac{r^2}{l^2}\,. \label{eq07}
\ee
Particularly, setting the symmetry breaking parameter $\eta=0$, this solution would reduce to standard four-dimensional Reissner-Nordstr\"{o}m AdS black hole.

Based on the line element (\ref{eq06}), we shall first calculate thermodynamical quantities of the black holes. To begin with, the event horizon of the black hole is located at $r=r_h$ which is the largest root of $f(r)=h(r)=0$. The electric charge and corresponding potential measured at infinity can be computed by
\bea
Q &=& \frac{1}{4 \pi} \oint F^{\mu \nu} d^2 \Sigma_{\mu \nu}=\left(1-\eta ^2\right)q\,, \\ \label{eq08}
\Phi &=& \xi^{\mu} A_{\mu} \big |_{r=r_h}=\frac{q}{r_h}\,, \label{eq09}
\eea
where $\xi^{\mu}$ is the timelike killing vector $(\partial _t)^{\mu}$ \cite{Wald:1984gr}. The Arnowitt-Deser-Misner (ADM) mass $M$ of the system can be determined via the Komar integral
\be
M=\frac{1}{8\pi}\oint \xi_{(t)}^{\mu;\nu }d^2\Sigma_{\mu \nu }=(1-\eta^2)m\,. \label{eq10}
\ee
The surface gravity for the black holes can be worked out from the relation \cite{Poisson:2004ar}
\be
\kappa^2=-\frac{1}{2}\xi_{\mu;\nu}\xi^{\mu;\nu} \big |_{r=r_h}\,. \label{eq11}
\ee
So the surface gravity is
\be
\kappa=\frac{1}{2}\sqrt{f_{,r} h_{,r}}~ \Big |_{r=r_h}=\frac{1}{2 r_h} \left[1+\frac{3 r_h^2}{l^2}-\frac{Q^2}{\left(1-\eta^2\right)^2 r_h^2}\right]\,, \label{eq12}
\ee
where we have expressed $\kappa$ as the function of $Q$ for convenience. And then, the Hawking temperature will be
\be
T=\frac{\kappa}{2 \pi}=\frac{1}{4 \pi r_h} \left[1+\frac{3 r_h^2}{l^2}-\frac{Q^2}{\left(1-\eta^2\right)^2 r_h^2}\right]\,. \label{eq13}
\ee
Moreover, the area $A_{bh}$ of the horizon and entropy $S$ can be obtained
\be
A_{bh}=\int_{r=r_h}\sqrt{g_{\theta \theta}g_{\phi \phi}}~d\theta d\phi=4\pi (1-\eta^2) r_h^2\,, \label{eq14}
\ee
\be
S=\frac{A_{bh}}{4}=\pi (1-\eta ^2) r_h^2\,. \label{eq15}
\ee
For an asymptotically AdS black hole in the extended phase space, the cosmological constant is usually interpreted as a thermodynamic pressure \cite{Kastor:2009wy}
\be
P=-\frac{\Lambda}{8 \pi}=\frac{3}{8 \pi l^2}\,. \label{eq16}
\ee
The corresponding conjugate thermodynamic volume $V=\left(\frac{\pt M}{\pt P}\right)_{S,Q_{i},J_{k}}$ is given by
\be
V=\frac{4}{3} \pi \left(1-\eta^2\right)r_h^3\,. \label{eq17}
\ee
Obviously, in the presence of a global monopole, thermodynamical quantities, including the electric charge $Q$, ADM mass $M$, the area $A_{bh}$ of the horizon and the volume $V$, Hawking temperature $T$ as well as entropy $S$, are closely related with parameter $\eta$. It can be verified that these quantities above still obey the first law of black hole thermodynamics
\be
dM=T dS + \Phi dQ + V dP\,, \label{eq18}
\ee
and the Smarr relation below holds as well
\be
M=2(T S-V P)+\Phi Q\,. \label{eq19}
\ee
It should be noted that here $M$ has had the meaning of the enthalpy $H$, but not the internal energy any longer \cite{Dolan:2012jh}.

Therefore, in comparison with the case for Reissner-Nordstr\"{o}m AdS black hole \cite{Kubiznak:2012wp}, the thermodynamical quantities of the black holes exhibit an interesting dependence on the internal global monopole. While, the existence of a global monopole does not affect the first law of thermodynamics and the Smarr relation.

\section{Phase transition and critical behaviors in the canonical ensemble} \label{section3}

With these thermodynamical quantities in hand, we step forward to further discuss the attractive properties of the charged AdS black holes with a global monopole, including phase transition and critical phenomena.

\subsection{The local stability and critical behaviors} \label{section3.1}

\begin{figure*}[htbp]
  \begin{center}
  \mbox{
    \subfigure[~$T$ vs. $r_h$ for varying charge]{\label{TRhdiag}\includegraphics[width=2.6in,height=2.3in]{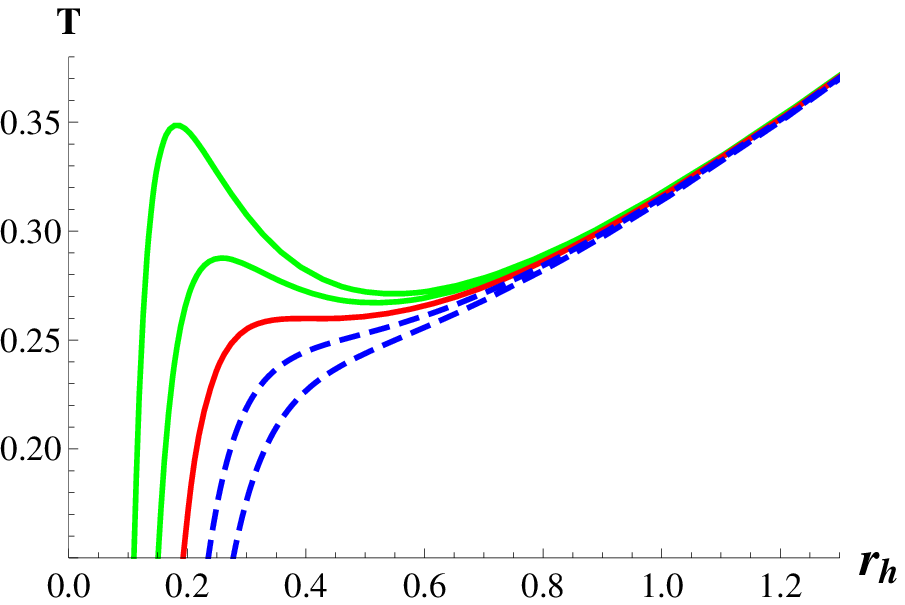}} \quad
    \subfigure[~$T$ vs. $S$ for varying charge]{\label{TSdiag}\includegraphics[width=2.6in,height=2.3in]{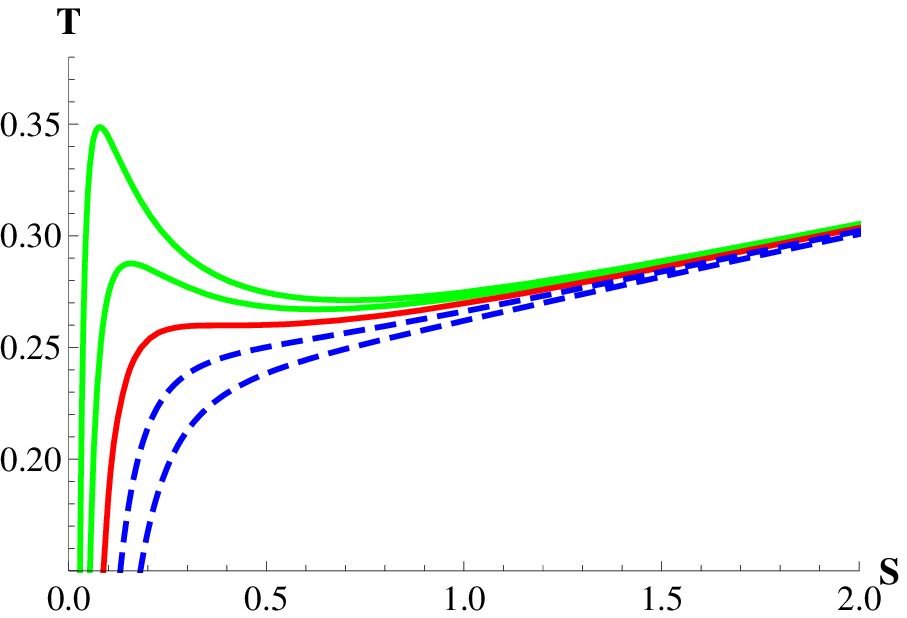}}
       }
    \caption{(Color online)\,. Behaviors of Hawking temperature. In each picture, the curves from top (green solid) to bottom (blue dashed) correspond to $Q=0.075,0.1,0.125,0.15,0.175$ respectively. Here we set $\eta=0.5,l=1$.}
  \label{TRhSdiag}
  \end{center}
\end{figure*}

The Hawking temperature is defined in equation (\ref{eq13}), its behaviors for different electric charges are depicted in Figure \ref{TRhSdiag}.
As is observed from both the Figure \ref{TRhdiag} and Figure \ref{TSdiag}, for each case of $Q<0.125$, within a certain range of temperature, the curve can be divided into three branches. The large and the small radius branches are thermodynamically stable, while the intermediate is unstable since the heat capacity $C=T\frac{\pt S}{\pt T}$ is negative. When the charge $Q\leq0.125$, the temperature could have an infection point for a given value of $Q$ and $l$. For the case $Q>0.125$, the Hawking temperature increases monotonically along each isocharge. Therefore, the analysis summarized in Figure \ref{TRhSdiag} might have an implication for phase transition.

In what follows, we focus on illustrating the phase structure and analyzing the critical behaviors of the black holes swallowing a global monopole in the canonical ensemble.

\begin{figure*}[htbp]
  \begin{center}
  \mbox{
    \subfigure[~$P$ vs. $\upsilon$ for the charged AdS black hole with a global monopole]{\label{Pvdiag1}\includegraphics[width=2.6in,height=2.3in]{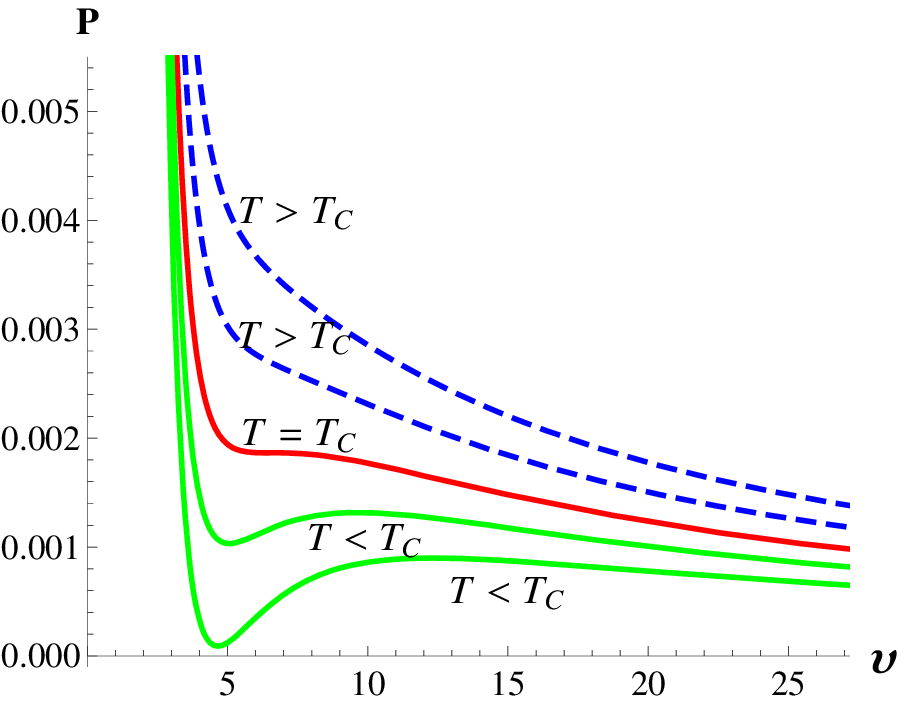}} \quad
    \subfigure[~$P$ vs. $\upsilon$ for Reissner-Nordstr\"{o}m AdS black holes]{\label{Pvdiag2}\includegraphics[width=2.6in,height=2.3in]{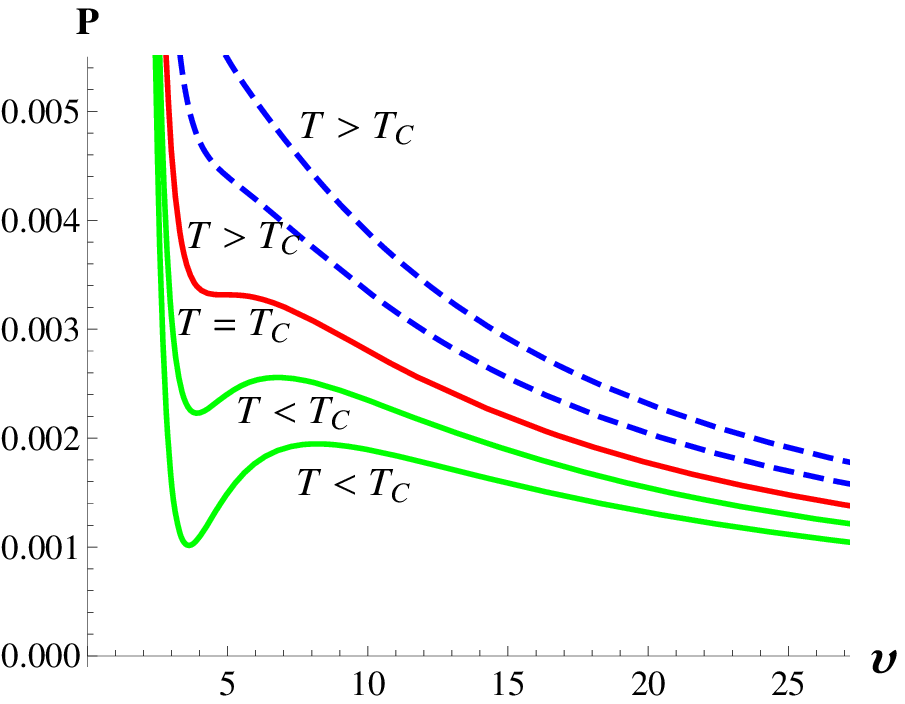}}
       }
    \caption{(Color online)\,. $P$-$\upsilon$ diagrams of the charged AdS black hole with a global monopole and Reissner-Nordstr\"{o}m AdS black holes. In each diagram, the temperature of isotherms decreases from top to bottom, and the middle red solid curve denotes the critical isotherm $T=T_c$. Here we set $\eta=0.5,Q=1$.}
  \label{Pvdiag}
  \end{center}
\end{figure*}

For a fixed electric charge $Q$, combing equations (\ref{eq13}) and (\ref{eq16}) yields the equation of state as
\be
P=\frac{T}{2 r_h}-\frac{1}{8 \pi r_h^2}+\frac{Q^2}{8 \pi \left(1-\eta ^2\right)^2 r_h^4}\,, \label{eq20}
\ee
where $r_h$ is a function of the thermodynamic volume (\ref{eq17}). To elaborate more physically, here we recast the geometric equation of state (\ref{eq20}) into a physical one by employing dimensional analysis as Ref.\cite{Kubiznak:2012wp}. The Planck length $l_P^2=G_N \hbar/c^3$, and the physical pressure, temperature can be given by
\be
\textit{Press}=\frac{\hbar c}{l_P^2}P\,,\quad \textit{Temp}=\frac{\hbar c}{k}T\,. \label{eq21}
\ee
Multiplying equation (\ref{eq20}) with $\frac{\hbar c}{l_P^2}$ and comparing with the Van der Waals equation \cite{Gunasekaran:2012dq}, we can obtain the relation of the specific volume $\upsilon$ with the horizon radius $r_h$ as
\be
\upsilon=2 l_P^2 r_h\,. \label{eq22}
\ee
Thus, the physical equation of state for the black hole system can be expressed as follows
\be
P=\frac{T}{\upsilon}-\frac{1}{2 \pi \upsilon^2}+\frac{2 Q^2}{\pi \left(1-\eta^2\right)^2 \upsilon^4}\,. \label{eq23}
\ee
Behaviors of the pressure can be witnessed intuitively in Figure \ref{Pvdiag1}.
It is evident that, similarly to the case of Reissner-Nordstr\"{o}m AdS black holes\footnote{The well-known case has been analyzed in Ref.\cite{Kubiznak:2012wp}. We display its $P$-$\upsilon$ diagram in Figure \ref{Pvdiag2}.}, the behaviors of pressure for the charged AdS black hole with a global monopole qualitatively behave like a Van der Waals liquid-gas system. There exists a small-large black hole phase transition in this system. When the temperature $T<T_c$, the isotherms are characterized by the intriguing Van der Waals oscillation. Each isotherm has a local minimum and maximum. Moreover, there is an infection point in each isotherm. With temperature increasing to $T=T_c$, the minimum and the maximum points approach to each other. Meanwhile, the oscillating segment squeezes to the critical point $(P_c,\upsilon_c,T_c)$. For the case $T>T_c$, the inflection point disappears, and the pressure decreases monotonically along each isotherm. Comparing Figure \ref{Pvdiag1} with Figure \ref{Pvdiag2}, it is quite interesting to note that the existence of a global monopole leads to a lower critical point. We can get insight into the reason for that by deriving the critical quantities. Utilizing equation (\ref{eq23}), one can explicitly determine the critical point via
\be
\left(\frac{\pt P}{\pt \upsilon}\right)_T = \left(\frac{\pt^2 P}{\pt \upsilon^2}\right)_T = 0\,, \label{eq24}
\ee
and get
\be
P_c=\frac{\left(1-\eta^2\right)^2}{96 \pi Q^2}\,,\quad \upsilon_c=\frac{2 \sqrt{6} Q}{1-\eta^2}\,,\quad T_c=\frac{1-\eta^2}{3 \sqrt{6} \pi Q}\,. \label{eq25}
\ee
These formulas indeed reflect that, in the presence of a global monopole, both of the critical pressure $P_c$ and the temperature $T_c$ decrease, while the critical volume $\upsilon_c$ increases, in contrast to the results of standard Reissner-Nordstr\"{o}m AdS black holes.

It's also interesting that the critical coefficient
\be
\frac{P_c \upsilon_c}{T_c}=\frac{3}{8}\,, \label{26}
\ee
is exactly in coincidence with Reissner-Nordstr\"{o}m AdS black holes although a global monopole is taken into account. Furthermore, performing the definitions
\be
p=\frac{P}{P_c}\,,\quad \nu=\frac{\upsilon}{\upsilon_c}\,,\quad \tau=\frac{T}{T_c}\,, \label{eq27}
\ee
the equation of state (\ref{eq23}) appears as a same law of corresponding states \cite{Kubiznak:2012wp}
\be
8\tau=3 \nu \left(p+\frac{2}{\nu^2}\right)-\frac{1}{\nu^3}\,, \label{eq28}
\ee
which agrees with the result in the mean field theory.

\begin{figure*}[htbp]
  \begin{center}
  \mbox{
    \subfigure[~$C_P$ vs. $r_h$ for $P=0.5P_c<P_c$]{\label{CpRhdiag0}\includegraphics[width=2.6in,height=2.3in]{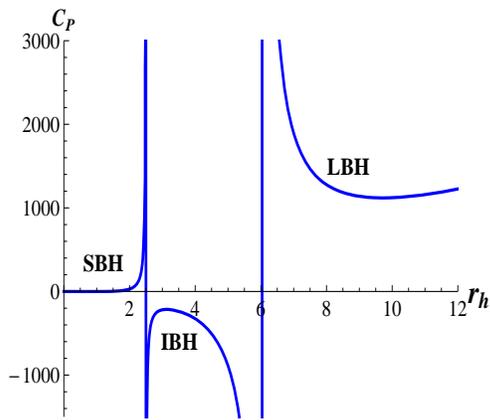}} \qquad
    \subfigure[~$C_P$ vs. $r_h$ for $P=P_c$]       {\label{CpRhdiag1}\includegraphics[width=2.6in,height=2.3in]{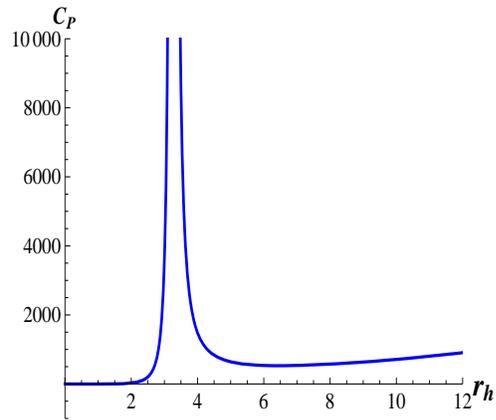}}
       }
  \mbox{
    \subfigure[~$C_P$ vs. $r_h$ for $P=2.8P_c>P_c$]{\label{CpRhdiag2}\includegraphics[width=2.6in,height=2.3in]{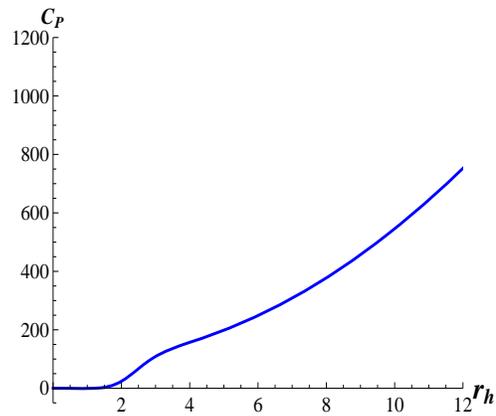}}
       }
    \caption{(Color online)\,. Heat capacity for varying pressure. Here we set $\eta=0.5,Q=1$.}
  \label{CpRhdiag}
  \end{center}
\end{figure*}

To further study the phase transition of the charged AdS black holes with a global monopole, we move to explore another important thermodynamical quantity, namely, the heat capacity $C_P$. In general, a positive heat capacity allows a stable black hole to exist, while a negative one signals that the black hole will disappear when suffering from a small perturbation. Considering equations (\ref{eq13}),(\ref{eq15}) and (\ref{eq16}), $C_P$ can be written as
\be
C_P=T \left(\frac{\pt S}{\pt T}\right)_P=\frac{2 \pi \left(1-\eta^2\right) r_h^2 \left[\left(1-\eta^2\right)^2 r_h^2 \left(8 \pi  P r_h^2+1\right)-Q^2\right]}{\left(1-\eta^2\right)^2 r_h^2 \left(8 \pi  P r_h^2-1\right)+3 Q^2}\,. \label{eq29}
\ee
We plot the heat capacity $C_P$ changing with $r_h$ in Figure \ref{CpRhdiag}.
Apparently, the curve of heat capacity for $P<P_c$ has two divergent points while that for $P=P_c$ has only one divergent point. When $P<P_c$, Figure \ref{CpRhdiag0} displays three regions divided by two divergent points. Both the small radius region and the large radius region are thermodynamically stable with positive heat capacity, while the intermediate radius region is unstable due to negative heat capacity. So the intermediate black hole (IBH) is unstable, and the phase transition takes place between the small black hole (SBH) and the large black hole (LBH). When $P>P_c$, as is shown in Figure \ref{CpRhdiag2}, the heat capacity always remains positive, implying that the black holes are locally stable and no phase transition will take place.

Up till now, we have analyzed the local thermodynamical properties of the charged AdS black hole incorporating a global monopole. It turns out that the black holes undergo a small-large black hole phase transition which is remarkably analogous to a Van der Waals liquid-gas system. Comparing this case with that of Reissner-Nordstr\"{o}m AdS black holes, we find that the global monopole affects the critical point while the critical coefficient and the law of corresponding states keep unchanged. In detail, in the presence of a global monopole, both of the critical pressure $P_c$ and the temperature $T_c$ decrease, while the critical volume $\upsilon_c$ increases. Next, we will proceed to illustrate the global stability by investigating Gibbs free energy.

\subsection{Gibbs free energy and the global stability} \label{section3.2}
\begin{figure*}[htbp]
  \begin{center}
  \mbox{
    \subfigure[~$G$ vs. $T$ for the charged AdS black hole with a global monopole]{\label{GTdiag0}\includegraphics[width=2.6in,height=2.3in]{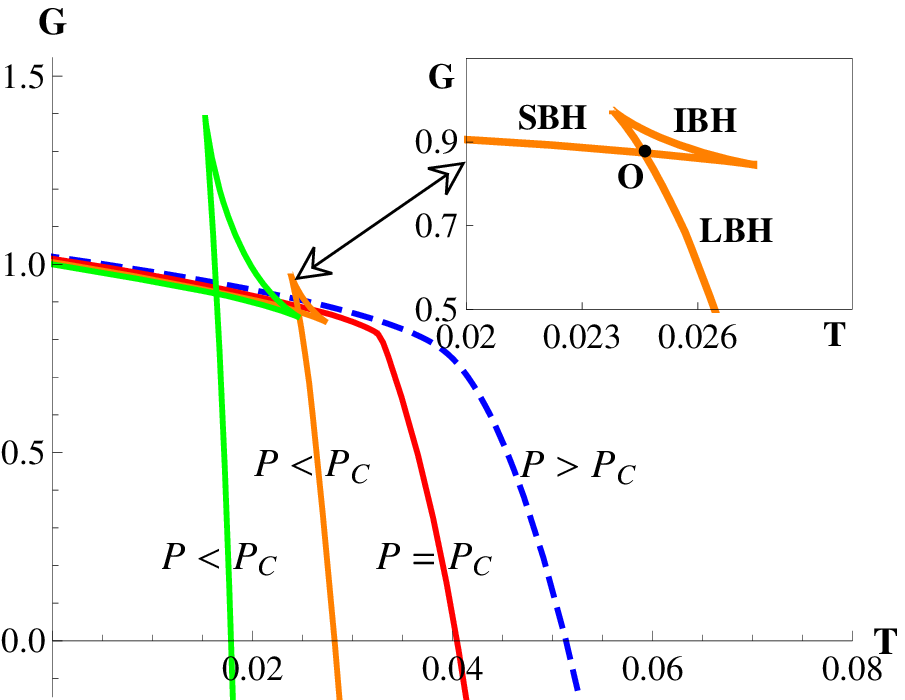}} \quad
    \subfigure[~$G$ vs. $T$ for Reissner-Nordstr\"{o}m AdS black holes]{\label{GTdiag1}\includegraphics[width=2.6in,height=2.3in]{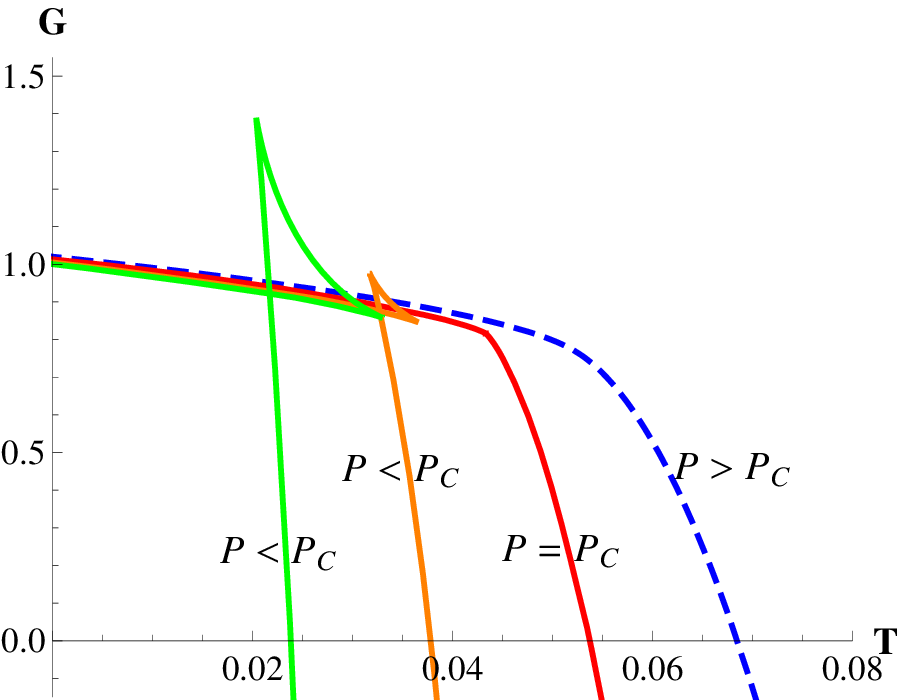}}
       }
    \caption{(Color online)\,. Behaviors of Gibbs free energy for the charged AdS black hole with a global monopole and Reissner-Nordstr\"{o}m AdS black holes. In each picture, the curves from right (blue dashed) to left (green solid) correspond to $P=1.6P_c,P_c,0.5P_c,0.2P_c$ respectively. Here we set $\eta=0.5,Q=1$.}
  \label{GTdiag}
  \end{center}
\end{figure*}

To facilitate discussing further, Gibbs free energy of the black hole system with a global monopole is derived as
\be
G=M-T S=\frac{1}{4}\left[\frac{3 Q^2}{\left(1-\eta^2\right) r_h}+\left(1-\eta^2\right) r_h (1-\frac{8 \pi}{3} P r_h^2)\right]\,, \label{eq30}
\ee
where $r_h$ is understood as a function of $P$ and $T$ via equation (\ref{eq20}). We plot the Gibbs free energy $G$ changing with $T$ for fixed Q in Figure \ref{GTdiag0}.
Seen from Figure \ref{GTdiag0}, with the pressure falling below the critical value $P_c$, the Gibbs free energy $G$ strikingly displays the characteristic swallow tail which indicates that the phase transition is of the first order. Observing the swallow tail more clearly (e.g., the isobar $P=0.5P_c$), the system first goes along the small black hole (SBH) branch until the temperature increases to the point $O$. At the point $O$, both the small and the large black holes share the same Gibbs free energy and can coexist. With the temperature increasing further, the large black hole (LBH) dramatically becomes the preferred thermodynamical state because of lower Gibbs free energy than the small and the intermediate black holes (IBH). Hence, there is a small-large black hole phase transition at the point $O$. What is more, due to the relation $S={A_{bh}}/4$, different horizon area for the small and the large black holes during the transitions corresponds to the discontinuity of the entropy. That means there is a release of latent heat, and the phase transition at the point $O$ is of the first order. In addition, comparing Figure \ref{GTdiag0} with Figure \ref{GTdiag1}, the effect of a global monopole on the Gibbs free energy of the black holes can be better understood, although they both clearly exhibit the classical swallow tail.

\section{Conclusions} \label{section4}
In this paper, we first calculated the thermodynamical quantities of the charged AdS black holes with a global monopole. The results show that, in the presence of a global monopole, thermodynamical quantities are closely related with the symmetry breaking parameter $\eta$. In contrast to the case of standard Reissner-Nordstr\"{o}m AdS black holes, they exhibit an interesting dependence on the internal global monopole. In spite of this, the quantities are verified to be still satisfy the first law of black hole thermodynamics and Smarr relation in the extended phase space. Moreover, we also analyzed the local and the global properties and revealed that the black hole undergoes a small-large phase transition. The critical behaviors and phase structure of the black holes were also discussed. We determined the critical point, the critical coefficient and the law of corresponding states. The critical point depends on the parameter $\eta$, while the two latters do not. More importantly, further illustrations manifest that the transition is of first order and remarkably analogous to a Van der Waals liquid-gas system. This paper not only contributes to deeper understanding black hole thermodynamics, but also gives an insight into the effect of a global monopole on the critical behaviors. Besides, it is of great importance to further support that Van der Waals-like behavior of the black holes is a universal phenomenon.

\begin{center}
{\bf Acknowledgements}
\end{center}
The authors would like to express sincere gratitude to the anonymous referee(s) for reviewing and their constructive suggestions which improve this paper greatly. We also appreciate the editor's time on this paper. This work is supported by the National Natural Science Foundation of China under Grant Nos. 11275017 and 11173028.


\begin{thebibliography}{99}

\bibitem{Barriola:1989hx}
M. Barriola, A. Vilenkin,
\textit{Phys. Rev. Lett.} {\bf 63}, 341 (1989).

\bibitem{Vilenkin:1994cs}
A. Vilenkin, E. P. S Shellard, \textit{Cosmic strings and other topological defects} (Cambridge University Press, New York, 2000).

\bibitem{Jing:1993np}
J. L. Jing, H. W. Yu, Y. J. Wang,
\textit{Phys. Lett. A} {\bf 178}, 59 (1993).

\bibitem{Yu:1994fy}
H. W. Yu,
\textit{Nucl. Phys. B} {\bf 430}, 427 (1994).

\bibitem{Dadhich:1997mh}
N. Dadhich, K. Narayan, U. A. Yajnik,
\textit{Pramana} {\bf 50}, 307 (1998),
arXiv:gr-qc/9703034.

\bibitem{Li:2002ku}
X. Z. Li, J. G. Hao,
\textit{Phys. Rev. D} {\bf 66}, 107701 (2002),
arXiv:hep-th/0210050.

\bibitem{Gao:2002hf}
C. J. Gao, Y. G. Shen,
\textit{Chin. Phys. Lett.} {\bf 19}, 477 (2002).

\bibitem{Jiang:2005xb}
Q. Q. Jiang, S. Q. Wu,
\textit{Phys. Lett. B} {\bf 635}, 151 (2006),
arXiv:hep-th/0511123.

\bibitem{Brihaye:2005qr}
Y. Brihaye, B. Hartmann, E.~Radu,
\textit{Phys. Rev. D} {\bf 74}, 025009 (2006),
arXiv:hep-th/0511305.

\bibitem{Han:2005qy}
Y. W. Han, S. Z. Yang, W. B. Liu,
\textit{Commun. Theor. Phys.} {\bf 43}, 382 (2005).

\bibitem{Wu:2007sw}
S. Q. Wu, J. J. Peng,
\textit{Class. Quant. Grav.} {\bf 24}, 5123 (2007),
arXiv:0706.0983 [hep-th].

\bibitem{Chen:2009vz}
S. B. Chen, L. C. Wang, C. K. Ding, J. J. Jing,
\textit{Nucl. Phys. B} {\bf 836}, 222 (2010),
arXiv:0912.2397 [gr-qc].

\bibitem{Lin:2011zzd}
K. Lin, J. Li, N. Yang,
\textit{Gen. Rel. Grav.} {\bf 43}, 1889 (2011).

\bibitem{Man:2013sf}
J. Y. Man, H. B. Cheng,
\textit{Phys. Rev. D} {\bf 87}, 044002 (2013),
arXiv:1301.2739 [hep-th].

\bibitem{Sharif:2015kna}
M. Sharif, S. Iftikhar,
\textit{Adv. High Energy Phys.} {\bf 2015}, 854264 (2015).

\bibitem{Ahmed:2016ucs}
A. K. Ahmed, U. Camci, M. Jamil,
\textit{Class. Quant. Grav.} {\bf 33}, 215012 (2016),
arXiv:1610.01129 [gr-qc].

\bibitem{Jusufi:2017lsl}
K. Jusufi, M. C. Werner, A. Banerjee, A.\"{o}vg\"{u}n,
\textit{Phys. Rev. D} {\bf 95}, 104012 (2017),
arXiv:1702.05600 [gr-qc].

\bibitem{Li:2017kze}
H. L. Li, S. R. Chen,
\textit{Gen. Rel. Grav.} {\bf 49}, 128 (2017).

\bibitem{Carames:2017ngt}
T. R. P. Caram\^{e}s, J. C. Fabris, E. R. Bezerra de Mello, H. Belich,
\textit{Eur. Phys. J. C} {\bf 77}, 496 (2017),
arXiv:1706.02782 [gr-qc].

\bibitem{Hawking:1982dh}
S. W. Hawking, D. N. Page,
\textit{Commun. Math. Phys.} {\bf 87}, 577 (1983).

\bibitem{Chamblin:1999tk}
A. Chamblin, R. Emparan, C. V. Johnson, R. C. Myers,
\textit{Phys. Rev. D} {\bf 60}, 064018 (1999).
arXiv:hep-th/9902170.

\bibitem{Chamblin:1999hg}
A. Chamblin, R. Emparan, C. V. Johnson, R. C. Myers,
\textit{Phys. Rev. D} {\bf 60}, 104026 (1999),
arXiv:hep-th/9904197.

\bibitem{Kubiznak:2012wp}
D. Kubiznak, R. B. Mann,
\textit{JHEP} {\bf 1207}, 033 (2012),
arXiv:1205.0559 [hep-th].

\bibitem{Kastor:2009wy}
D. Kastor, S. Ray, J. Traschen,
\textit{Class. Quant. Grav.} {\bf 26}, 195011 (2009),
arXiv:0904.2765 [hep-th].

\bibitem{Cvetic:2010jb}
M. Cvetic, G. W. Gibbons, D. Kubiznak, C. N. Pope,
\textit{Phys. Rev. D} {\bf 84}, 024037 (2011),
arXiv:1012.2888 [hep-th].

\bibitem{Dolan:2012jh}
B. P. Dolan,
arXiv:1209.1272 [gr-qc].

\bibitem{Caceres:2015vsa}
E. Caceres, P. H. Nguyen, J. F. Pedraza,
\textit{JHEP} {\bf 1509}, 184 (2015),
arXiv:1507.06069 [hep-th].

\bibitem{Gunasekaran:2012dq}
S. Gunasekaran, R. B. Mann, D. Kubiznak,
\textit{JHEP} {\bf 1211}, 110 (2012),
arXiv:1208.6251 [hep-th].

\bibitem{Belhaj:2012bg}
A. Belhaj, M. Chabab, H. El Moumni, M. B. Sedra,
\textit{Chin. Phys. Lett.} {\bf 29}, 100401 (2012),
arXiv:1210.4617 [hep-th].

\bibitem{Cai:2013qga}
R. G. Cai, L. M. Cao, L. Li, R. Q. Yang,
\textit{JHEP} {\bf 1309}, 005 (2013),
arXiv:1306.6233 [gr-qc].

\bibitem{Zhao:2013oza}
R. Zhao, H. H. Zhao, M. S. Ma, L. C. Zhang,
\textit{Eur. Phys. J. C} {\bf 73}, 2645 (2013),
arXiv:1305.3725 [gr-qc].

\bibitem{Ma:2013aqa}
M. S. Ma, H. H. Zhao, L. C. Zhang and R. Zhao,
\textit{Int. J. Mod. Phys. A} {\bf 29}, 1450050 (2014),
arXiv:1312.0731 [hep-th].

\bibitem{Zou:2013owa}
D. C. Zou, S. J. Zhang, B. Wang,
\textit{Phys. Rev. D} {\bf 89}, 044002 (2014),
arXiv:1311.7299 [hep-th].

\bibitem{Dehghani:2014caa}
M. H. Dehghani, S. Kamrani, A. Sheykhi,
\textit{Phys. Rev. D} {\bf 90}, 104020 (2014),
arXiv:1505.02386 [hep-th].

\bibitem{Xu:2014kwa}
W. Xu and L. Zhao,
\textit{Phys. Lett. B} {\bf 736}, 214 (2014),
arXiv:1405.7665 [gr-qc].

\bibitem{Johnson:2014pwa}
C. V. Johnson,
\textit{Class. Quant. Grav.} {\bf 31}, 225005 (2014),
arXiv:1406.4533 [hep-th].

\bibitem{Wei:2014hba}
S. W. Wei, Y. X. Liu,
\textit{Phys. Rev. D} {\bf 90}, 044057 (2014),
arXiv:1402.2837 [hep-th].

\bibitem{Wei:2015iwa}
S. W. Wei and Y. X. Liu,
\textit{Phys. Rev. Lett.} {\bf 115}, 111302 (2015),
arXiv:1502.00386 [gr-qc].

\bibitem{Hendi:2014kha}
S. H. Hendi, S. Panahiyan, B. Eslam Panah,
\textit{Int. J. Mod. Phys. D} {\bf 25}, 1650010 (2015),
arXiv:1410.0352 [gr-qc].

\bibitem{Guo:2015waa}
X. Y. Guo, H. F. Li, L. C. Zhang, R. Zhao,
\textit{Phys. Rev. D} {\bf 91}, 084009 (2015).

\bibitem{Mo:2016sel}
J. X. Mo, G. Q. Li,
\textit{Phys. Rev. D} {\bf 92}, 024055 (2015),
arXiv:1604.07931 [gr-qc].

\bibitem{Sherkatghanad:2014hda}
Z. Sherkatghanad, B. Mirza, Z. Mirzaiyan and S. A. Hosseini Mansoori,
\textit{Int. J. Mod. Phys. D} {\bf 26}, 1750017 (2016),
arXiv:1412.5028 [gr-qc].

\bibitem{Fernando:2016sps}
S. Fernando,
\textit{Phys. Rev. D} {\bf 94}, 124049 (2016),
arXiv:1611.05329 [gr-qc].

\bibitem{Hendi:2016njy}
S. H. Hendi, S. Panahiyan, B. Eslam Panah, M. Faizal, M. Momennia,
\textit{Phys. Rev. D} {\bf 94}, 024028 (2016),
arXiv:1607.06663 [gr-qc].

\bibitem{Pradhan:2016dun}
P. Pradhan,
arXiv:1603.07750 [gr-qc].

\bibitem{Zeng:2016aly}
S. He, L. F. Li, X. X. Zeng,
\textit{Nucl. Phys. B} {\bf 915}, 243 (2017),
arXiv:1608.04208 [hep-th].

\bibitem{Poshteh:2016rwc}
M. B. J. Poshteh, N. Riazi,
\textit{Gen. Rel. Grav.} {\bf 49}, 64 (2017),
arXiv:1606.01408 [hep-th].

\bibitem{Bhattacharya:2017hfj}
K. Bhattacharya, B. R. Majhi,
\textit{Phys. Rev. D} {\bf 95}, 104024 (2017),
arXiv:1702.07174 [gr-qc].

\bibitem{Wald:1984gr}
R. M. Wald, \textit{General relativity} (University of Chicago Press, Chicago, 1984).

\bibitem{Poisson:2004ar}
E. Poisson, \textit{A relativist's toolkit: The mathematics of black-hole mechanics} (Cambridge University Press, New York, 2004).

\end{thebibliography}
\end{document}